\documentclass[prl,twocolumn,showpacs,floatfix,amsmath,amssymb,superscriptaddress,longbibliography]{revtex4-2}
\usepackage{amsfonts}
\usepackage{stmaryrd}
\usepackage{bbm}
\usepackage{mathrsfs}
\usepackage{tipa}
\usepackage{amssymb}
\usepackage{txfonts}
\usepackage{graphicx}
\usepackage{dcolumn}
\usepackage{epstopdf}
\usepackage[colorlinks,linkcolor=blue,urlcolor=black,citecolor=blue]{hyperref}
\usepackage{multirow}
\usepackage{subfigure}
\usepackage{url}
\usepackage{upgreek}
\usepackage[utf8]{inputenc}
\usepackage[english]{babel}
\usepackage{siunitx}

\begin{document}

\title{\Large Nonthermal nonlinear many-body excitonic dynamics in high-purity silicon}

\author{Patrick Pilch}
\author{Siyu Duan}
\affiliation{ 
Department of Physics, TU Dortmund University, 44227 Dortmund, Germany
}

\author{Yang Liu}
\affiliation{ 
Center for Correlated Matter and School of Physics, Zhejiang University, Hangzhou, China
}
\affiliation{ 
Institute for Advanced Study in Physics, Zhejiang University, Hangzhou 310058, China
}

\author{Manfred Bayer}
\affiliation{ 
Department of Physics, TU Dortmund University, 44227 Dortmund, Germany
}
\affiliation{Research Center FEMS, TU Dortmund University, 44227 Dortmund, Germany}
\affiliation{
Dortmund Center for Advanced Exploration of Dynamics Across Limits Using Spectroscopy (DAEDALUS), TU Dortmund University, 44227 Dortmund, Germany
}

\author{Marc Assmann}
\author{Zhe~Wang}
\affiliation{ 
Department of Physics, TU Dortmund University, 44227 Dortmund, Germany
}
\affiliation{
Dortmund Center for Advanced Exploration of Dynamics Across Limits Using Spectroscopy (DAEDALUS), TU Dortmund University, 44227 Dortmund, Germany
}

\date{\today}

\begin{abstract}
We report time-resolved optical pump-terahertz third-harmonic generation spectroscopy of picosecond nonthermal many-body charge dynamics in high-purity silicon at room temperature.
Although in thermal equilibrium excitons in silicon are unstable due to a lower binding energy than the thermal energy at room temperature,
following pulsed non-resonant optical pumping we can identify the formation of transient excitonic states on the picosecond time scale in our high-resolution nonlinear spectroscopic measurements.
In contrast to unbound electron-hole pairs, the nonthermal many-body excitonic dynamics feature a characteristic time-dependent evolution, a distinct dependence on charge density, and a nonperturbative nonlinear terahertz response.
\end{abstract}

\maketitle


An exciton is a bosonic composite quasiparticle of an electron-hole pair bound by attractive Coulomb interaction.
The bosonic character and many-body interactions of excitons make them fundamentally appealing for the realization of exotic quantum phenomena in solid-state materials, e.g. Bose-Einstein condensation \cite{Butov1994,Nat.Commu.2022observation}, exciton fluids \cite{delAguila2023}, excitonic insulators \cite{Cercellier07,Ma2021} or exciton crystals \cite{Qi2026}.
The formation and transport dynamics of excitons are crucial also for applications, such as semiconductor optoelectronics or quantum technologies \cite{Su2008,Wilson2021}.
Experimental studies of excitonic dynamics have mainly been based on spectroscopic methods focusing on the linear-response regime.
For example, photoluminescence spectroscopy measures emitted photons due to recombination of the electron and hole in an exciton \cite{Butov1994,Nat.Commu.2022observation,delAguila2023};
Absorption or reflection spectroscopy detects the coherent excitonic polarization induced by optical excitation \cite{Nature2014giant,Nat.Commu.2022observation,Qi2026};
Also incoherent excitons can be probed by their electromagnetic field-driven transport properties particularly in the terahertz (1~THz = 10$^{12}$~Hz) frequency range \cite{Huber2001,kira2001exciton, Kaindl2003,koch2006semiconductor}, while a \textit{dc} field cannot induce a current of the charge-neutral objects.

In this work, by using a different spectroscopic technique that goes beyond the linear response and thermal equilibrium -- time-resolved optical pump THz high-harmonic generation spectroscopy -- we investigate many-body excitonic dynamics in the indirect bandgap semiconductor silicon at room temperature.
Whereas without optical excitation the high-purity silicon is transparent for THz electromagnetic waves, with optical excitation of electron-hole pairs and strong THz drive pulses we are able to detect THz third-harmonic generation (THG), which corresponds to the lowest order nonzero nonlinearity since second-order nonlinear responses vanish due to inversion symmetry of silicon.
In this optically-excited strongly interacting many-body system, complex dynamical processes are taking place simultaneously.
The observation of the high-order nonlinear response allows us to sensitively track the formation or dissociation of excitons down to picosecond time scales. 
In particular, we resolve characteristic THz third-harmonic generation of the far-from-equilibrium picosecond excitonic dynamics by their distinctive dependence on charge carrier density featuring a Mott crossover and their nonperturbative dependence on THz drive strength.

Charge dynamics in silicon plays a key role in its widespread applications, ranging from computers and smart phones, to lasers, LEDs, and solar cells.
Characterized by an indirect band gap of about 1.12~eV at room temperature, phonon-assisted excitation of electron-hole pairs in silicon can be triggered by visible or near-infrared photons.
The binding energy of excitons in silicon is about 14.7~meV \cite{ShakleeNahory1970}, corresponding to a thermal energy of $T=170$~K.
Therefore, in thermal equilibrium, excitons in silicon are not stable at room temperature due to strong thermal fluctuations.
Hence, previous THz spectroscopic studies of excitonic dynamics in silicon were performed at liquid-helium temperatures where excitons are thermally stable and a coherent intraexcitonic absorption line of the quasi-equilibrium excitons is well resolvable \cite{Suzuki09,SuzukiPRB2011,SuzukiPRL2012}.
In contrast to these previous studies, here we study a very different regime of excitonic dynamics, i.e. far-from-equilibrium picosecond ultrafast nonlinear dynamics, which is distinct on the following aspects:
Firstly, we focus on room temperature where the lifetime of excitons is expected to be shorter due to enhanced thermal fluctuations compared with cryogenic temperatures.
Secondly, following the optical creation of electron-hole pairs, with strong THz pulse we drive the nonthermal many-body system even further away from thermal equilibrium.
Thirdly, we employ the time-resolved nonlinear (rather than linear) response to sense many-body dynamics of the strongly interacting system.


\begin{figure*}[t!]
\centering 
\includegraphics[width=0.9\linewidth]{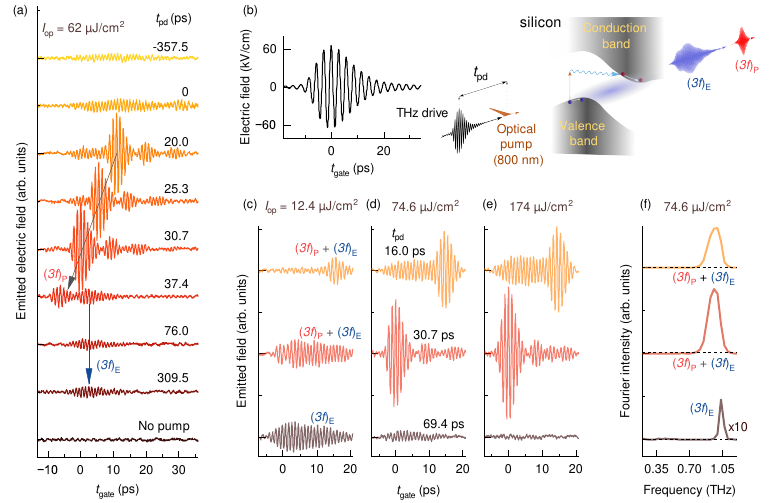}
\caption{(a) THz third harmonic amplitudes for various pump-drive delays $t_{\mathrm{pd}}$ at a fixed optical pump fluence of $I_\text{op}=62~\upmu$J/cm$^2$ and a driving THz electric field of 65~kV/cm.
With varying $t_{\mathrm{pd}}$, a shifting and a non-shifting third-harmonic signal are resolved as indicated by the arrows and denoted by  $(3f)_\mathrm{P}$ and $(3f)_\mathrm{E}$ for plasma and excitonic emission, respectively.
(b) Waveform of the 0.35~THz drive pulse, and schematics of THz field-driven charge carrier dynamics after phonon-assisted optical excitations in the indirect bandgap semiconductor silicon.
The valence and conduction bands are characterized by nonparabolic dispersion relations, which is crucial for efficient nonlinear THz emission \cite{Madelung02}. 
Emitted third-harmonic electric field in the time domain measured at three well-separated pump-drive delays $t_{\mathrm{pd}}$ for optical pump fluences of (c) $I_\text{op}=12.4$, (d) 74.6, and (e) 174~$\upmu$J/cm$^2$.
(f) The corresponding frequency-domain spectra for $I_\text{op}=74.6~\upmu$J/cm$^2$.
}
\label{Fig1} 
\end{figure*}

For our spectroscopic study, we use an intrinsic high-resistive $\langle 100\rangle$-oriented silicon wafer with resistivity $>10^5$~$\Omega$cm and a thickness of 0.5~mm.
To excite electron-hole pairs across the indirect gap, we use 1.55~eV photons \cite{Suzuki09,SuzukiPRB2011,SuzukiPRL2012,MORIYASU2024130139, LI20124102} from a pulsed laser system (800~nm) with a pulse duration of 100~fs and a repetition rate of 1~kHz.
Since silicon has a direct band gap of 3.43~eV, electron-hole pairs across this gap cannot be directly excited by 1.55~eV photons through either linear absorption or second-order nonlinear optical processes.
Based on the same laser system, our intense multicycle THz drive pulses are obtained by bandpass filtering THz radiation from optical rectification in a LiNbO$_3$ crystal \cite{zhu2024population,ZhuPRL2025}.
For an $f=0.35$~THz drive pulse [see Fig.~\ref{Fig1}(b) for its waveform], we used two 0.35~THz bandpass filters each with $\pm26$\% bandwidth at $-20$~dB.
With respect to the THz drive, the arrival time of the optical pump pulse at the sample position is tunable, with the corresponding pump-drive time difference denoted by $t_\text{pd}$ [see Fig.~\ref{Fig1}(b) for illustration].
$t_\text{pd}$ is defined as positive when the pump arrives before the drive pulse. 
The THz electric field is determined by electro-optic sampling with a ZnTe crystal \cite{wu1995free}, in which the time delay of a gating optical pulse with respect to the THz pulse is denoted by $t_\text{gate}$.
The measurements are performed at room temperature under ambient conditions.
 
Under the drive of an $f=0.35$~THz pulse with a peak electric field of 65~kV/cm, the emitted third-harmonic radiation is recorded after two 1~THz bandpass filters each with $\pm30$\% bandwidth at $-20$~dB.
The obtained time-domain traces are presented in Fig.~\ref{Fig1}(a) for various pump-drive delays and an optical pump fluence of 62~$\upmu$J/cm$^2$. 
For contrast, the trace recorded without optical pump displays no THz third-harmonic oscillations, which confirms the high purity of the intrinsic silicon sample, since doped silicon samples are not transparent in the THz frequencies and emit THz third-harmonic radiation even without optical pump \cite{MengPRB2020,MengPRB2022,MengPRR2023}. 
A strong THG yield appears in intrinsic silicon after the optical excitation, reflecting THz driven nonlinear charge dynamics of the optically excited states.
As indicated by the two arrows in Fig.~\ref{Fig1}(a), two third-harmonic traces denoted by $(3f)_\mathrm{P}$ and $(3f)_\mathrm{E}$, respectively, can be distinguished by their different dependencies on the pump-drive delay $t_{\mathrm{pd}}$.
While the $(3f)_\mathrm{P}$ traces shift in the time domain following the shifting optical pump with respect to the THz drive pulse, the $(3f)_\mathrm{E}$ traces remain at the same position. 
Besides their very different dependence on the pump-drive delay, as will be discussed below, the two traces exhibit a rich set of distinctly different spectral characteristics.  
These features allow us to assign the $(3f)_\mathrm{P}$ and $(3f)_\mathrm{E}$ traces as plasma and excitonic THG, respectively.

\begin{figure}[t]
\centering 
\includegraphics[width=0.95\linewidth]{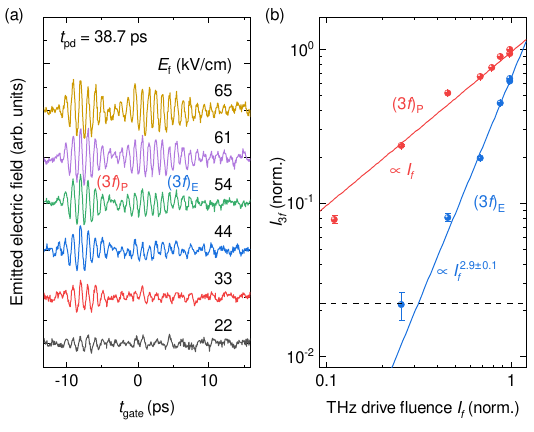}
\caption{Time-domain third-harmonic traces measured with $I_{\mathrm{op}}=74.6~\upmu$J/cm$^2$ for different THz drive fields from 22 to 65~kV/cm
(a) at $t_{\mathrm{pd}}$~=~38.7~ps where two third-harmonic contributions $(3f)_E$ and $(3f)_P$ are well separated. 
(b) The integrated third-harmonic intensity $I_{3f}$ of the plasma contribution exhibits a linear dependence on the THz drive fluence $I_{f}$, while for the exciton contribution $I_{3f} \propto I^{2.9\pm 0.1}_{f}$.
The noise level is indicated by the horizontal dashed lines.}
\label{Fig:THz_Fluence} 
\end{figure}

It is worth emphasizing that the excitonic third-harmonic emission can be distinguished from the plasma THG with a temporal separation even shorter than 8~ps [see e.g. the $t_{\mathrm{pd}}=37.4$~ps curve in Fig.~\ref{Fig1}(a)], whereas from a linear-response probe \cite{Suzuki09} a coherent intraexcitonic 1\textit{S}-2\textit{P} excitation became resolvable at least 50~ps after the optical excitation.
This indicates that excitonic correlations are built up from the electron-hole plasma on a shorter time scale due to the Coulomb interaction which is sensitively probed in our nonlinear spectroscopy, while a fully bound electron-hole pair is formed at later stage which is manifested by the resonant absorption in linear THz probe spectroscopy \cite{KiraKoch2006}. The rapid building up of excitonic correlations relative to the THz drive-pulse duration of approximately 30~ps [Fig.~\ref{Fig1}(b)] and the relatively long lifetime of excitons \cite{Suzuki09} allow the detection of the temporally separated plasma and excitonic THG for a single pump-drive delay at $t_\text{pd} \gtrsim 30$~ps [e.g. $t_{\mathrm{pd}}=37.4$~ps in Fig.~\ref{Fig1}(a)].
Moreover, no excitonic second-harmonic generation was measurable under the same conditions, which is consistent with the centrosymmetric structure of silicon and excludes substantial nonlinear effects due to symmetry breaking at the surfaces or field-induced symmetry breaking.

In addition, we observe a weak third-harmonic signal at negative pump-drive delays, see e.g. $t_\text{pd} = -357.5$~ps in Fig.~\ref{Fig1}(a), in contrast to the intuition that any THz third-harmonic yield should appear after the optical pump.
This signal corresponds to the dynamics of the residual photoexcited electrons and holes from the preceding pulse, which is one millisecond earlier. 
Since under quasi-thermal equilibrium conditions fully bound excitons cannot survive on millisecond time scales at room temperature, the observed slow dynamics may be most intuitively attributed to trapped carriers due to possible imperfection at the sample surface. 
Dynamics on such slow time scales in this indirect bandgap semiconductor \cite{hammond1980temperature} has been subject of previous studies by using time-resolved linear THz spectroscopy \cite{Caddle23,MORIYASU2024130139, LI20124102}. In contrast, we focus on picosecond ultrafast dynamics in this work.

The resolved distinct differences in the time-domain THG traces demonstrate the powerfulness of our time-resolved high-harmonic generation spectroscopy, which enables us to identify and investigate coexisting dynamical processes from unbound plasma or bound excitonic pairs in the strongly interacting nonthermal many-body system.
While the nonparabolic dispersion relation in the conduction and valence bands \cite{Madelung02} can lead to the third-harmonic yields of the THz driven charge dynamics, our results show that the driven plasma and excitonic dynamics behave very differently on several important aspects. 

Firstly, we study the different nonlinear responses of the THz driven plasma and excitonic dynamics by varying the THz field strength. 
In Fig.~\ref{Fig:THz_Fluence}(a), we present the results for $t_{\mathrm{pd}}$~=~38.7~ps, where the excitonic THG traces $(3f)_\mathrm{E}$ are behind and clearly separated from the plasma THG $(3f)_\mathrm{P}$ in the time domain.
With increasing THz drive fluence, the excitonic THG increases much more evidently than the plasma THG.
By performing Fourier analysis of the corresponding time-domain traces separately, we obtain plasma and excitonic THG intensity spectra in the frequency domain, from which we derive integrated THG intensity $I_{3f}$ versus THz drive fluence $I_f$ as displayed in a log-log scale in Fig.~\ref{Fig:THz_Fluence}(b).
As shown by the solid lines, the excitonic THG exhibits a nearly perturbative cubic power-law dependence of $I_{3f} \propto I^{2.9\pm 0.1}_{f}$, whereas the plasma THG follows a linear dependence on the THz drive. 
The linear dependence of the plasma THG is attributed to the photoinduced change in the THz optical conductivity due to the photoexcitation of charge carriers and their intraband dynamics, while the nonlinear dependence of excitonic THG arises naturally from band nonparabolicity.

\begin{figure}[t!]
\centering 
\includegraphics[width=0.9\linewidth]{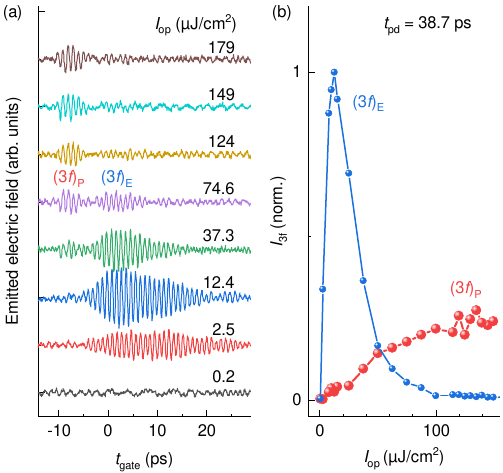}
\caption{(a) Optical pump fluence dependence of third-harmonic yields  measured with a THz drive field of 65~kV/cm at a pump-drive delay of $t_\text{pd}=38.7$~ps, where two THG contributions are resolvable and labelled by $(3f)_\mathrm{P}$ and $(3f)_\mathrm{E}$. 
(b) Integrated THG intensity $I_{3f}$ versus optical pump fluence $I_\text{op}$, as derived from the time-domain signals in (a) for the two contributions $(3f)_\mathrm{E}$ and $(3f)_\mathrm{P}$ separately.}
\label{Fig:PumpFluence}
\end{figure}

Secondly, with a fixed THz drive field of 65~kV/cm, we study the evolution of the THz THG by changing the optical pump fluence $I_\text{op}$ (see Fig.~\ref{Fig:PumpFluence}). 
We have found distinctly different pump-fluence dependencies for the excitonic and the plasma THG.
For $t_\text{pd}=38.7$~ps, a THz third-harmonic yield is hardly resolvable at $I_\text{op}=0.2~\upmu$J/cm$^2$ [Fig.~\ref{Fig:PumpFluence}(a)], but already very evident with a relatively low pump fluence of $I_\text{op}=2.5~\upmu$J/cm$^2$.
At low fluences, the increase of the excitonic THG yield with enhanced pump fluence is very rapid [see also Fig.~\ref{Fig:PumpFluence}(b)], whereas the observed plasma THG remains weak.
The maximum excitonic THG is reached around $I_\text{op}=12.4~\upmu$J/cm$^2$, which is followed by a continuous decrease towards the highest fluences.
In contrast, the plasma THG increases monotonically and tends to saturate above about $I_\text{op}=100~\upmu$J/cm$^2$.

The integrated THG intensity of the plasma and excitonic dynamics is plotted separately in Fig.~\ref{Fig:PumpFluence}(b) as a function of the optical pump fluence. 
The distinct difference in their optical pump dependencies confirms the assignment of the two THG traces as plasma and excitonic THG, respectively.
The observed behaviors can be understood as follows.
At low pump fluences corresponding to a low charge density, the initial increase of charge density caused by higher pump fluences leads to the formation of more excitons, hence enhancing both the plasma and excitonic third-harmonic yields.
At the same time, the interactions between charges and excitons are increasing with charge density.
Above a critical charge density, dissociation of excitons occurs due to screening mediated by the enhanced Coulomb interaction. 
They decay into unbound electron-hole pairs and contribute to the plasma, and eventually excitons cannot be formed in the presence of a large electron-hole density, i.e., in a Mott phase \cite{SuzukiPRL2012,Shah77}.
Therefore, above a critical density a decrease of the excitonic THG is observed experimentally [Fig.~\ref{Fig:PumpFluence}(b)], while the plasma THG increases monotonically towards saturation.

From the excitation pulse energy corresponding to the strongest excitonic THG observed at $I_\text{op}=12.4~\upmu$J/cm$^2$, we can estimate the photoexcited electron-hole pair density as $n=\SI{3.3e16}{\per\cubic\centi\meter}$, while the excitonic THG essentially vanishes beyond $I_\text{op}=120~\upmu$J/cm$^2$ corresponding to $n\gtrsim\SI{3e17}{\per\cubic\centi\meter}$.
Just for comparison, the theoretical Mott density of $n_{\mathrm{Mott}}=\SI{4.4e17}{\per\cubic\centi\meter}$ at room temperature under thermal equilibrium is not much greater, which is given by assuming extrapolation of the known theoretical results \cite{NorrisBajaj82,Rogers70} to room temperature.
However, due to strong thermal fluctuations the excitons in silicon cannot survive at room temperature under thermal equilibrium conditions. 
Hence, the observed carrier-density-dependent behaviors here should be understood as transient dynamical properties of the strongly interacting many-body electron-hole system far from thermal equilibrium.
 
\begin{figure}[t]
\centering 
\includegraphics[width=1\linewidth]{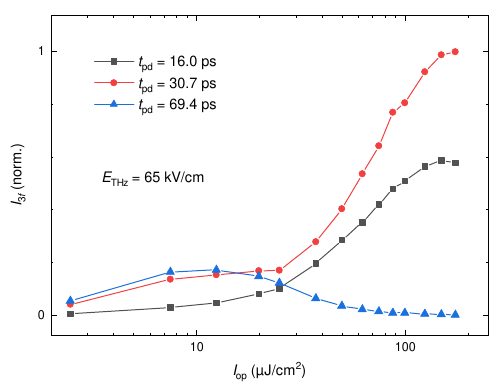}
\caption{Optical pump fluence dependence of the overall third-harmonic intensity measured with a peak THz electric field of 65~kV/cm for various pump-drive delays.}
\label{Fig:tPD} 
\end{figure}

Thirdly, at different pump-drive delays, the overall THz THG intensity exhibits evidently different dependence on the optical pump fluence (see Fig.~\ref{Fig:tPD}), reflecting the distinct difference of plasma and excitonic dynamics. 
For a shorter pump-drive delay, e.g. $t_\text{pd}=16.0$~ps, where a large density of excitons has not been formed, the observed third-harmonic yield is governed by plasma dynamics, therefore a monotonic increase of THG due to increased charge-carrier density at a higher pump fluence is observed [see Fig.~\ref{Fig1}(c)-(e) and Fig.~\ref{Fig:tPD}].  
In contrast, at a rather long pump-drive delay of $t_\text{pd}=69.4$~ps, where the optical pump pulse is well ahead of the THz drive pulse [see Fig.~\ref{Fig1}(a)], the observed third-harmonic yield is mainly due to the driven excitonic dynamics [see Fig.~\ref{Fig1}(f)], leading to a Mott-type nonmonotonic dependence of THG intensity (Fig.~\ref{Fig:tPD}).
At low fluence the THG intensity increases due to larger density of excitons, but above the critical value of $12.4~\upmu$J/cm$^2$ the third-harmonic yield decreases monotonically due to the suppression of the picosecond exciton formation by enhanced Coulomb interactions [c.f. the $t_\text{pd}=69.4$~ps curves in Fig.~\ref{Fig1}(c)-(e)].
For an intermediate pump-drive delay (see $t_\text{pd}=30.7$~ps in Fig.~\ref{Fig:tPD}), the third-harmonic yield at low pump fluences is characterized by excitonic dynamics as featured by a broad maximum, whereas at high pump fluences dominated by plasma contributions.

In conclusion, by using time-resolved optical pump THz third-harmonic generation spectroscopy, we have investigated picosecond nonthermal nonlinear many-body dynamics of an optically excited strongly interacting electron-hole system in high-purity silicon at room temperature.
We have shown that under the drive of an intense narrowband THz pulse, both the plasma and excitonic dynamics lead to the emission of THz radiation at the third-harmonic frequency.
For plasma, this corresponds to a broadband radiation dominated by a linear THz field dependence, which is dominant when the pump and drive pulses overlap.
In contrast, the driven excitonic third-harmonic emission is narrowband and characterized by a nonperturbative power-law THz field dependence.
Moreover, with increasing charge-carrier density the plasma emission increases monotonically, while the excitonic third-harmonic generation exhibits a maximum at a critical density, above which strong Coulomb interactions lead to dissociation of the excitons, featuring a Mott crossover behavior.


\begin{acknowledgements}
We acknowledge support by the European Research Council (ERC) under the Horizon 2020 research and innovation programme, Grant Agreement No. 950560 (DynaQuanta).
S.D. was supported by the Walter Benjamin Programme of the German Research Foundation (DFG) via the Project Number 557566396.
Y.L. was supported by the National Science Foundation of China (Grants No. 12525408).
\end{acknowledgements}

\bibliography{Silicon_bib}

@article{Rogers70,
  title = {Bound Eigenstates of the Static Screened Coulomb Potential},
  author = {Rogers, F. J. and Graboske, H. C. and Harwood, D. J.},
  journal = {Phys. Rev. A},
  volume = {1},
  issue = {6},
  pages = {1577--1586},
  numpages = {0},
  year = {1970},
  month = {Jun},
  publisher = {American Physical Society},
  doi = {10.1103/PhysRevA.1.1577},
  url = {https://link.aps.org/doi/10.1103/PhysRevA.1.1577}
}

@article{NorrisBajaj82,
  title = {{Exciton-plasma Mott transition in Si}},
  author = {Norris, George B. and Bajaj, K. K.},
  journal = {Phys. Rev. B},
  volume = {26},
  issue = {12},
  pages = {6706--6710},
  numpages = {0},
  year = {1982},
  month = {Dec},
  publisher = {American Physical Society},
  doi = {10.1103/PhysRevB.26.6706},
  url = {https://link.aps.org/doi/10.1103/PhysRevB.26.6706}
}

@article{KiraKoch2006,
title = {Many-body correlations and excitonic effects in semiconductor spectroscopy},
journal = {Progress in Quantum Electronics},
volume = {30},
number = {5},
pages = {155-296},
year = {2006},
issn = {0079-6727},
doi = {https://doi.org/10.1016/j.pquantelec.2006.12.002},
url = {https://www.sciencedirect.com/science/article/pii/S0079672706000280},
author = {M. Kira and S.W. Koch}
}

@article{Caddle23,
    author = {Butler-Caddle, Edward and Grant, Nicholas E. and Pain, Sophie L. and Murphy, John D. and Jayawardena, K. D. G. Imalka and Lloyd-Hughes, James},
    title = {Terahertz photoconductance dynamics of semiconductors from sub-nanosecond to millisecond timescales},
    journal = {Applied Physics Letters},
    volume = {122},
    number = {1},
    pages = {012101},
    year = {2023},
    month = {01},
    issn = {0003-6951},
    doi = {10.1063/5.0130721},
    url = {https://doi.org/10.1063/5.0130721},
    eprint = {},
}

@article{ShakleeNahory1970,
  title = {{Valley-Orbit Splitting of Free Excitons? The Absorption Edge of Si}},
  author = {Shaklee, K. L. and Nahory, R. E.},
  journal = {Phys. Rev. Lett.},
  volume = {24},
  issue = {17},
  pages = {942--945},
  numpages = {0},
  year = {1970},
  month = {Apr},
  publisher = {American Physical Society},
  doi = {10.1103/PhysRevLett.24.942},
  url = {https://link.aps.org/doi/10.1103/PhysRevLett.24.942}
}

@Article{delAguila2023,
author={del {\'A}guila, Andr{\'e}s Granados and Wong, Yi Ren and Wadgaonkar, Indrajit and Fieramosca, Antonio and Liu, Xue and Vaklinova, Kristina and Dal Forno, Stefano and Do, T. Thu Ha and Wei, Ho Yi and Watanabe, K. and Taniguchi, T. and Novoselov, Kostya S. and Koperski, Maciej and Battiato, Marco and Xiong, Qihua},
title={{Ultrafast exciton fluid flow in an atomically thin MoS$_2$ semiconductor}},
journal={Nat. Nanotechnol.},
year={2023},
month={Sep},
day={01},
volume={18},
number={9},
pages={1012-1019},
issn={1748-3395},
doi={10.1038/s41565-023-01438-8},
url={https://doi.org/10.1038/s41565-023-01438-8}
}

@article{Cercellier07,
  title = {{Evidence for an Excitonic Insulator Phase in $1T\mathrm{\text{\ensuremath{-}}}{\mathrm{TiSe}}_{2}$}},
  author = {Cercellier, H. and Monney, C. and Clerc, F. and Battaglia, C. and Despont, L. and Garnier, M. G. and Beck, H. and Aebi, P. and Patthey, L. and Berger, H. and Forr\'o, L.},
  journal = {Phys. Rev. Lett.},
  volume = {99},
  issue = {14},
  pages = {146403},
  numpages = {4},
  year = {2007},
  month = {Oct},
  publisher = {American Physical Society},
  doi = {10.1103/PhysRevLett.99.146403},
  url = {https://link.aps.org/doi/10.1103/PhysRevLett.99.146403}
}

@article{Butov1994,
  title = {{Condensation of Indirect Excitons in Coupled AlAs/GaAs Quantum Wells}},
  author = {Butov, L. V. and Zrenner, A. and Abstreiter, G. and B\"ohm, G. and Weimann, G.},
  journal = {Phys. Rev. Lett.},
  volume = {73},
  issue = {2},
  pages = {304--307},
  numpages = {0},
  year = {1994},
  month = {Jul},
  publisher = {American Physical Society},
  doi = {10.1103/PhysRevLett.73.304},
  url = {https://link.aps.org/doi/10.1103/PhysRevLett.73.304}
}

@article{Su2008,
author={Su, Jung-Jung
and MacDonald, A. H.},
title={How to make a bilayer exciton condensate flow},
journal={Nat. Phys.},
year={2008},
month={Oct},
day={01},
volume={4},
number={10},
pages={799-802},
issn={1745-2481},
doi={10.1038/nphys1055},
url={https://doi.org/10.1038/nphys1055}
}

@article{Qi2026,
author={Qi, Ruishi
and Li, Qize
and Kim, Haleem
and Nie, Jiahui
and Zhang, Zuocheng
and Xia, Ruichen
and Cui, Zhiyuan
and Xiao, Jianghan
and Taniguchi, Takashi
and Watanabe, Kenji
and Crommie, Michael F.
and Wang, Feng},
title={An exciton crystal in a moir{\'e} excitonic insulator},
journal={Nat. Phys.},
year={2026},
month={Feb},
day={18},
url={https://doi.org/10.1038/s41567-026-03184-9}
}

@Article{Ma2021,
author={Ma, Liguo
and Nguyen, Phuong X.
and Wang, Zefang
and Zeng, Yongxin
and Watanabe, Kenji
and Taniguchi, Takashi
and MacDonald, Allan H.
and Mak, Kin Fai
and Shan, Jie},
title={Strongly correlated excitonic insulator in atomic double layers},
journal={Nature},
year={2021},
month={Oct},
day={01},
volume={598},
number={7882},
pages={585-589},
issn={1476-4687},
doi={10.1038/s41586-021-03947-9},
url={https://doi.org/10.1038/s41586-021-03947-9}
}

@article{Shah77,
  title = {{Investigation of Exciton-Plasma Mott Transition in Si}},
  author = {Shah, Jagdeep and Combescot, M. and Dayem, A. H.},
  journal = {Phys. Rev. Lett.},
  volume = {38},
  issue = {25},
  pages = {1497--1500},
  numpages = {0},
  year = {1977},
  month = {Jun},
  publisher = {American Physical Society},
  doi = {10.1103/PhysRevLett.38.1497},
  url = {https://link.aps.org/doi/10.1103/PhysRevLett.38.1497}
}

@article{Suzuki09,
  title = {{Time-Resolved Formation of Excitons and Electron-Hole Droplets in {S}i Studied Using Terahertz Spectroscopy}},
  author = {Suzuki, Takeshi and Shimano, Ryo},
  journal = {Phys. Rev. Lett.},
  volume = {103},
  issue = {5},
  pages = {057401},
  numpages = {4},
  year = {2009},
  month = {Jul},
  publisher = {American Physical Society},
  doi = {10.1103/PhysRevLett.103.057401},
  url = {https://link.aps.org/doi/10.1103/PhysRevLett.103.057401}
}

@Article{Kaindl2003,
author={Kaindl, R. A.
and Carnahan, M. A.
and H{\"a}gele, D.
and L{\"o}venich, R.
and Chemla, D. S.},
title={Ultrafast terahertz probes of transient conducting and insulating phases in an electron--hole gas},
journal={Nature},
year={2003},
month={Jun},
day={01},
volume={423},
number={6941},
pages={734-738},
issn={1476-4687},
doi={10.1038/nature01676},
url={https://doi.org/10.1038/nature01676}
}

@Article{Huber2001,
author={Huber, R.
and Tauser, F.
and Brodschelm, A.
and Bichler, M.
and Abstreiter, G.
and Leitenstorfer, A.},
title={How many-particle interactions develop after ultrafast excitation of an electron--hole plasma},
journal={Nature},
year={2001},
month={Nov},
day={01},
volume={414},
number={6861},
pages={286-289},
issn={1476-4687},
doi={10.1038/35104522},
url={https://doi.org/10.1038/35104522}
}

@Article{Wilson2021,
author={Wilson, Nathan P.
and Yao, Wang
and Shan, Jie
and Xu, Xiaodong},
title={Excitons and emergent quantum phenomena in stacked 2{D} semiconductors},
journal={Nature},
year={2021},
month={Nov},
day={01},
volume={599},
number={7885},
pages={383-392},
issn={1476-4687},
doi={10.1038/s41586-021-03979-1},
url={https://doi.org/10.1038/s41586-021-03979-1}
}

@article{wu1995free,
  title={Free-space electro-optic sampling of terahertz beams},
  author={Wu, Qi and Zhang, X-C},
  journal={Appl. Phys. Lett.},
  volume={67},
  number={24},
  pages={3523--3525},
  year={1995},
  publisher={American Institute of Physics},
  url={https://ui.adsabs.harvard.edu/link_gateway/1995ApPhL..67.3523W/doi:10.1063/1.114909}
}

@article{Nature2014giant,
  title={Giant {R}ydberg excitons in the copper oxide {C}u$_2${O}},
  author={Kazimierczuk, Tomasz and Fr{\"o}hlich, Dietmar and Scheel, Stefan and Stolz, Heinrich and Bayer, Manfred},
  journal={Nature},
  volume={514},
  number={7522},
  pages={343--347},
  year={2014},
  publisher={Nature Publishing Group UK London},
  url={https://doi.org/10.1038/nature13832}
}

@article{Nat.Commu.2022observation,
  title={Observation of {B}ose-{E}instein condensates of excitons in a bulk semiconductor},
  author={Morita, Yusuke and Yoshioka, Kosuke and Kuwata-Gonokami, Makoto},
  journal={Nat. Commun.},
  volume={13},
  number={1},
  pages={5388},
  year={2022},
  publisher={Nature Publishing Group UK London},
  url={https://doi.org/10.1038/s41467-022-33103-4}
}

@article{zhu2024population,
  title = {Population inversion and ultrafast terahertz nonlinearity of transient Dirac fermions in {C}d$_3${A}s$_{2}$},
  author = {Zhu, Changqing and Pilch, Patrick and Dantas, Renato M. A. and Reinold, Anneke and Yang, Yunkun and Xiu, Faxian and Bedoya-Pinto, Amilcar and Parkin, Stuart S. P. and Moessner, Roderich and Wang, Zhe},
  journal = {Phys. Rev. B},
  volume = {109},
  issue = {20},
  pages = {L201111},
  numpages = {6},
  year = {2024},
  month = {May},
  publisher = {American Physical Society},
  doi = {10.1103/PhysRevB.109.L201111},
  url = {https://link.aps.org/doi/10.1103/PhysRevB.109.L201111}
}

@article{ZhuPRL2025,
  title = {Distinct Terahertz Third-Harmonic Generation of Many-Body Excitonic States},
  author = {Zhu, Changqing and Reinold, Anneke and Pilch, Patrick and Kovalev, Sergey and Heck\"otter, Julian and Reiter, Doris and Bayer, Manfred and Assmann, Marc and Wang, Zhe},
  journal = {Phys. Rev. Lett.},
  volume = {134},
  issue = {19},
  pages = {196503},
  numpages = {6},
  year = {2025},
  month = {May},
  publisher = {American Physical Society},
  doi = {10.1103/PhysRevLett.134.196503},
  url = {https://link.aps.org/doi/10.1103/PhysRevLett.134.196503}
}

@article{LI20124102,
title = {Photocarriers dynamics in silicon wafer studied with optical-pump terahertz-probe spectroscopy},
journal = {Opt. Commun.},
volume = {285},
number = {20},
pages = {4102-4106},
year = {2012},
issn = {0030-4018},
doi = {https://doi.org/10.1016/j.optcom.2012.05.053},
url = {https://www.sciencedirect.com/science/article/pii/S0030401812004853},
author = {Gaofang Li and Dong Li and Zuanming Jin and Guohong Ma}
}

@article{SuzukiPRB2011,
  title = {{Cooling dynamics of photoexcited carriers in Si studied using optical pump and terahertz probe spectroscopy}},
  author = {Suzuki, Takeshi and Shimano, Ryo},
  journal = {Phys. Rev. B},
  volume = {83},
  issue = {8},
  pages = {085207},
  numpages = {8},
  year = {2011},
  month = {Feb},
  publisher = {American Physical Society},
  doi = {10.1103/PhysRevB.83.085207},
  url = {https://link.aps.org/doi/10.1103/PhysRevB.83.085207}
}

@article{SuzukiPRL2012,
  title = {{Exciton Mott Transition in Si Revealed by Terahertz Spectroscopy}},
  author = {Suzuki, Takeshi and Shimano, Ryo},
  journal = {Phys. Rev. Lett.},
  volume = {109},
  issue = {4},
  pages = {046402},
  numpages = {5},
  year = {2012},
  month = {Jul},
  publisher = {American Physical Society},
  doi = {10.1103/PhysRevLett.109.046402},
  url = {https://link.aps.org/doi/10.1103/PhysRevLett.109.046402}
}

@article{MengPRB2020,
  title = {{Intracavity third-harmonic generation in Si:B pumped by intense terahertz pulses}},
  author = {Meng, Fanqi and Thomson, Mark D. and ul-Islam, Qamar and Klug, Bernhard and Pashkin, Alexej and Schneider, Harald and Roskos, Hartmut G.},
  journal = {Phys. Rev. B},
  volume = {102},
  issue = {7},
  pages = {075205},
  numpages = {9},
  year = {2020},
  month = {Aug},
  publisher = {American Physical Society},
  doi = {10.1103/PhysRevB.102.075205},
  url = {https://link.aps.org/doi/10.1103/PhysRevB.102.075205}
}

@article{MengPRB2022,
  title = {{Importance of valence-band anharmonicity and carrier distribution for third- and fifth-harmonic generation in Si:B pumped with intense terahertz pulses}},
  author = {Meng, Fanqi and Walla, Frederik and ul-Islam, Qamar and Pashkin, Alexej and Schneider, Harald and Jungemann, Christoph and Thomson, Mark D. and Roskos, Hartmut G.},
  journal = {Phys. Rev. B},
  volume = {106},
  issue = {7},
  pages = {075203},
  numpages = {7},
  year = {2022},
  month = {Aug},
  publisher = {American Physical Society},
  doi = {10.1103/PhysRevB.106.075203},
  url = {https://link.aps.org/doi/10.1103/PhysRevB.106.075203}
}

@article{MengPRR2023,
  title = {Higher-harmonic generation in boron-doped silicon from band carriers and bound-dopant photoionization},
  author = {Meng, Fanqi and Walla, Frederik and Kovalev, Sergey and Deinert, Jan-Christoph and Ilyakov, Igor and Chen, Min and Ponomaryov, Alexey and Pavlov, Sergey G. and H\"ubers, Heinz-Wilhelm and Abrosimov, Nikolay V. and Jungemann, Christoph and Roskos, Hartmut G. and Thomson, Mark D.},
  journal = {Phys. Rev. Res.},
  volume = {5},
  issue = {4},
  pages = {043141},
  numpages = {21},
  year = {2023},
  month = {Nov},
  publisher = {American Physical Society},
  doi = {10.1103/PhysRevResearch.5.043141},
  url = {https://link.aps.org/doi/10.1103/PhysRevResearch.5.043141}
}

@article{MORIYASU2024130139,
title = {{Photocarrier dynamics in thick Si film studied by optical pump-terahertz probe spectroscopy}},
journal = {Opt. Commun.},
volume = {554},
pages = {130139},
year = {2024},
issn = {0030-4018},
doi = {https://doi.org/10.1016/j.optcom.2023.130139},
url = {https://www.sciencedirect.com/science/article/pii/S0030401823008878},
author = {Takeshi Moriyasu and Masahiko Tani and Hideaki Kitahara and Takashi Furuya and Jessica Afalla and Toshiro Kohmoto and Daishiro Koide and Hiroki Sato and Mitsutaka Kumakura}
}

@article{koch2006semiconductor,
  title={Semiconductor excitons in new light},
  author={Koch, Stephan W and Kira, Mackillo and Khitrova, Galina and Gibbs, Hyatt M},
  journal={Nat. Mater.},
  volume={5},
  number={7},
  pages={523--531},
  year={2006},
  publisher={Nature Publishing Group UK London},
  url={https://doi.org/10.1038/nmat1658}
}

@article{kira2001exciton,
  title = {Exciton Formation in Semiconductors and the Influence of a Photonic Environment},
  author = {Kira, M. and Hoyer, W. and Stroucken, T. and Koch, S. W.},
  journal = {Phys. Rev. Lett.},
  volume = {87},
  issue = {17},
  pages = {176401},
  numpages = {4},
  year = {2001},
  month = {Oct},
  publisher = {American Physical Society},
  doi = {10.1103/PhysRevLett.87.176401},
  url = {https://link.aps.org/doi/10.1103/PhysRevLett.87.176401}
}

@article{hammond1980temperature,
  title={Temperature dependence of the exciton lifetime in high-purity silicon},
  author={Hammond, RB and Silver, RN},
  journal={Appl. Phys. Lett.},
  volume={36},
  number={1},
  pages={68--71},
  year={1980},
  publisher={AIP Publishing},
  url={https://doi.org/10.1063/1.91277}
}

@book{Madelung02,
  title     = {Landolt-Börnstein - Group III Condensed Matter
Volume 41A1$\beta$ (Group IV Elements, IV-IV and III-V Compounds. Part b - Electronic, Transport, Optical and Other Properties).},
  author    = {Madelung, O. and Rössler, U. and Schulz, M.},
  year      = {2002},
  publisher = {Springer}
}

\end{document}